\documentstyle[12pt]{article}

\def\be{\begin{equation}}
\def\ee{\end{equation}}
\def\bea{\begin{eqnarray}}
\def\eea{\end{eqnarray}}
\def\ben{\begin{enumerate}}
\def\een{\end{enumerate}}

\def\nnu{\nonumber}
\def\ll{\label}

\begin{document}
\newcommand{\Psl}{\not\!\! P}
\newcommand{\dsl}{\not\! \partial}
\newcommand{\half}{{\textstyle\frac{1}{2}}}
\newcommand{\for}{{\textstyle\frac{1}{4}}}
\newcommand{\eqn}[1]{(\ref{#1})}
\newcommand{\npb}[3]{ {\bf Nucl. Phys. B}{#1} ({#2}) {#3}}
\newcommand{\pr}[3]{ {\bf Phys. Rep. }{#1} ({#2}) {#3}}
\newcommand{\prl}[3]{ {\bf Phys. Rev. Lett. }{#1} ({#2}) {#3}}
\newcommand{\plb}[3]{ {\bf Phys. Lett. B}{#1} ({#2}) {#3}}
\newcommand{\prd}[3]{ {\bf Phys. Rev. D}{#1} ({#2}) {#3}}
\newcommand{\hepth}[1]{ [{\bf hep-th}/{#1}]}
\newcommand{\grqc}[1]{ [{\bf gr-qc}/{#1}]}

\def\la{\mathrel{\mathpalette\fun <}}
\def\a{\alpha}
\def\b{\beta}
\def\g{\gamma}\def\G{\Gamma}
\def\d{\delta}\def\D{\Delta}
\def\ep{\epsilon}
\def\et{\eta}
\def\z{\zeta}
\def\t{\theta}\def\T{\Theta}
\def\l{\lambda}\def\L{\Lambda}
\def\m{\mu}
\def\f{\phi}\def\F{\Phi}
\def\n{\nu}
\def\p{\psi}\def\P{\Psi}
\def\r{\rho}
\def\s{\sigma}\def\S{\Sigma}
\def\ta{\tau}
\def\x{\chi}
\def\o{\omega}\def\O{\Omega}
\def\k{\kappa}
\def\pa {\partial}
\def\ov{\over}
\def\br{\nonumber\\}
\begin{flushright}
DFPD/98/TH/03\\ hep-th/9801038\\Revised v2\\
\end{flushright}
\bigskip\bigskip
\begin{center}
{\large\bf NEW SUPERSYMMETRIC VACUA FOR D=4, N=4 GAUGED SUPERGRAVITY}
\footnote{ Research supported by INFN fellowship}

\vskip .9 cm
{\sc Harvendra Singh}
\footnote{e-mail: hsingh@pd.infn.it} 
 \vskip 0.05cm
INFN Sezione di Padova, Departimento di Fisica `Galileo Galilei', \\
Via F. Marzolo 8, 35131 Padova, Italy
\end{center}
\bigskip
\centerline{\bf ABSTRACT}
\bigskip

\begin{quote}
In this paper we obtain  supersymmetric brane-like 
configurations in the vacuum of 
$N=4$ gauged $SU(2)\times SU(2)$ supergravity theory in four 
spacetime dimensions. Almost all of
 these vacuum solutions preserve either half
or one quarter of the supersymmetry in the theory. We also study
these solutions in presence of nontrivial axionic and gauge field
backgrounds. In the case of pure
gravity with axionic charge the geometry of the spacetime is
$AdS_3\times R^1$ with $N=1$ supersymmetry.
An interesting observation is that the domain walls of this theory cannot 
be given an interpretation of a 2-brane in four dimensions. But it still  
exists as a stable vacuum. This feature  
is quite distinct from the domain-wall configuration in
massive type IIA supergravity in ten dimensions.

\end{quote}

\newpage
\section{Introduction}
Recent understanding of the nonperturbative aspects in string theories and 
its duality symmetries \cite{dev} has triggered a series of new 
developments in the study of massive supergravity theories in 
various dimensions
\cite{roma,berg,cow,ms,bre,hs,bct1,hlw, blo}. 
It is viewed that the massive supergravities are
required for the spacetime interpretations of higher dimensional world volume 
Dirichlet objects (D-p-branes) of type II superstrings. 
For example, massive (gauged) type IIA supergravity \cite{roma} is 
a candidate theory for type IIA 
D-8-brane \cite{pw, berg, bg} 
whose field strength is a 10-form. The dual of  a 10-form
in ten dimensions is a constant which can be identified as a mass term 
(cosmological constant) of massive
type IIA supergravity. Next, the strong coupling limit of type IIA strings
is M-thoery whose spacetime theory is sought to be the 
eleven dimensional supergravity \cite{cjs}.
String dualities require that all even branes of type IIA and odd branes of
IIB should have an 
eleven-dimensional interpretation \cite{tow}. In this picture a
D-8-brane can be obtained 
from a double dimensional reduction of M-9-brane or toroidal reduction of an
M-8-brane.  Therefore 11-dimensional supergravity should also
have massive analog of type IIA supergravity. 
There has been some progress recently in finding
 massive  11-dimensional supergravity 
\cite{hlw,blo}.
 
There are various ways to get massive supergravities starting from the
massless ones. For example the generalised Scherk and Schwarz dimensional 
reduction scheme 
\cite{ss} do give rise to massive supergravities. Under generalised reduction
scheme some of the higher dimensional fields are given linear dependences
along the directions to be compactified \cite{berg,cow}. This procedure
 has been used in 
\cite{berg} to demonstrate the type IIA and IIB duality in D=9 for the massive
case. Other methods require the gauging of internal global symmetries
in the supergravity theory \cite{dfr,fs}. 
An interesting  feature of  massive or gauged supergravity theories is that, in
general,  
they possess original amount of supersymmetry even in the presence of mass
(gauge) parameters. 
If supersymmetry is only the criterion then massive supergravities
are as consistent as massless ones.
However, distinction appears in terms of internal symmetries.
The price we pay is that the internal global
symmetries including dualities are  generally reduced to 
smaller symmetry groups during the process of gauging.  
Once the mass (or gauge)  parameters are 
set to zero   the `standard' (massless) supergravities are recovered. 

In this work we will study  $N=4$ gauged  $SU(2)\times SU(2)$ 
supergravity in four dimensions \cite{fs}. 
It has been recently established that this 
gauged supergravity can be embedded into $N=1$ supergravity in 
ten dimensions as an $S^3\times S^3$ compactification \cite{cv1} 
\footnote{I thank  Prof. P.K. Townsend 
bringing this fact into my knowledge}.
Previously also, a Kaluza-Klein(KK) 
interpretation for the $SU(2)\times SU(2)$ gauged 
supergravity was given in \cite{abs} where this model was identified as part of 
the effective $D=4$ field theory for the heterotic string theory in an
$S^3 \times S^3$ vacuum.   
These two KK interpretations are essentially the 
same upto consistent truncations. 
Thus it is worthwhile to study above $N=4, D=4$ gauged supergravity
keeping in mind the recent
developments in massive supergravities. 

It is well known that the dimensional reduction of ten-dimensional 
supersymmetric Yang-Mills theory to 4 dimensions gives us $SU(4)$ invariant
supersymmetric theory\cite{gso}. An off-shell supergravity with $SU(4)$
global symmetry  and $N=4$ local supersymmetry  was 
constructed in four dimensions long ago \cite{csf}. 
The spectrum of this theory  in the gravity 
multiplet consists; the graviton $E_\m^M$, 4 Majorana spin-$3/2$ gravitinos 
$\Psi_\m^I$, 3 vector fields $A_\m^a$, 3  axial-vector fields $B_\m^a$, 
4 spin-$1/2$
Majorana fields $\chi^I$, the dilaton field $\phi$, 
and a pseudo-scalar field $\eta $.
There do exist other versions of $N=4$ supergravity in $D=4$, e.g. $SO(4)$ 
supergravity theory\cite{das}. 
It turns out that at the equation of motion 
level  both the above supergravities are equivalent upto certain field 
redefinitions and duality relations. 
The internal subgroup $SU(2)\times SU(2)$  of $SU(4)$ supergravity 
can be gauged, giving rise to $N=4$  supergravity with local
$SU(2)\times SU(2)$ 
symmetry \cite{fs}. 
However, imposition of local supersymmetry in presence of the 
local internal symmetry requires introduction of  mass 
like terms (or cosmological constant) in the theory in addition to certain
bilinear fermionic terms \cite{fs}.  
The mass term appears as an effective dilaton potential 
in the theory.  
It was noted \cite{fs} 
that the dilaton potential leads to energy density which 
is unbounded from below which is  physically nonacceptable
\footnote{ Similar unbounded potential also arises in gauged
$SO(4)$ supergravity theory \cite{dfr}}. 
But the question can  be asked
whether this theory yields some stable vacuum configurations. 
It has been
noted that theory allows supersymmetric solutions like `electro-vac'
solutions\cite{fg} and more recently non-Abelian solitons as stable vacuum 
configurations were also shown to exist\cite{cv,cv1}.   
In related work on strings in curved backgrounds \cite{afk},
exact supersymmetric solutions of 
four-dimensional gauged supergravities have been constructed using the
powerful techniques of (super)conformal field theory by exploiting a
connection between gauged supergravities and non-critical strings.

In this work we will show that extended objects  like strings,  domain walls,
pure axionic gravity, dilaton-axion gravity
and maximally symmetric point-like configurations do exist in the vacuum
of the $N=4, D=4$ gauged theory. 
All these backgrounds preserve half or one-fourth of the
supersymmetries except for the point-like configuration for which
we have not been  able to find supersymmetric case. We also present the
generalisation of the fundamental string solution as a vacuum solution of the
massive supergravity.
The paper has been
organised as follows. In the next section we briefly describe the gauged model.
 In the subsequent sections III, IV, and V 
we obtain various background solutions and discuss there supersymmetric
properties in each case. In the last section VI we summarise our results.

\section{ The Gauged Model}
We consider the truncated version of the $N=4$ gauged $SU(2)_A\times SU(2)_B$ 
supergravity model \cite{fs} given  by the condition  
$B_\m^a=0$, that is half of the gauge fields are vanishing. In addition 
we shall only study the configurations which are either purely electric type
or magnetic type with respect to remaining 2-form field strength. 
We also set all the spinor fields to zero in our study. 
Under these specifications the  action \cite{fs} becomes
\bea
&& S= \int d^4x \sqrt{-g}~\bigg[~ R + {1\ov 2} \left(\pa_\m \f\pa^\m\f 
+ e^{2\f} \pa_\m \eta\pa^\m \eta \right) 
- {1\ov 2.2!} e^{-\f}F_{\m\n}^a F^{a\m\n}\nnu\\  
&&\hskip3cm - {1\ov 2.2!} ~ \eta~ F_{\m\n}^a \tilde F^{a\m\n}  
+{\L^2 \ov 2} e^\f \bigg]
\label{act}
\eea
where $\L^2=e_A^2+e_B^2$, $e_A$ and $e_B$ being the gauge couplings of 
respective $SU(2)$ groups and 
$$F_{\m\n}^a=\pa_\m A_\n^a -\pa_\n A_\m^a +e_A\epsilon_{a b c} A_\m^bA_\n^c,$$
$$ \tilde F_{\m\n}^a={1\ov 2} \ep_{\m\n}^{~~\s\r} F_{\s\r}^a.$$

The corresponding supersymmetry transformations are \cite{fs},
\bea
&&\delta\bar\chi^I={i\ov 2\sqrt{2}} \bar\ep^I\left(\pa_\m\f 
+i \g_5 e^\f\pa_\m\eta\right)\g^\m -
{1\ov 4} e^{-{\f\ov 2}}
\bar\ep^I\a^a F_{\m\n}^a \G^{\m\n} + {1\ov 4} e^{\f\ov 2}\bar\ep^I~
(e_A+i\g_5e_B),\nnu\\  
&&\delta\bar\Psi_\m^I=\bar\ep^I\left(\overleftarrow\pa_\m -{1\ov 2} 
\omega_{\m,m n}\G^{m n}
+{1\ov 2} e_A\a^a A_\m^a\right) -{i\ov 4} e^\f \bar\ep\g_5\pa_\m\eta
\nnu\\ &&\hskip1cm -{i\ov  4\sqrt{2}} e^{-{\f\ov 2}}\bar\ep^I
\a^a F_{\n\r}^a \gamma_\m \G^{\n\r} +{i\ov 4\sqrt{2}} e^{\f\ov 2} 
\bar\ep^I (e_A+i\g_5 e_B)\g_\m,
\nnu\\
&&\delta (bosons)= 0,
\label{2}
\eea
where $\ep^I$ are four spacetime dependent Majorana spinors. Since fermionic
backgrounds are absent therefore  the bosonic
fields do not vary under  supersymmetry. 
Our convention for the metric is with mostly minus signs $(+---)$, 
$\o_\m,^{m n}$ are the spin connections, 
$\{\g_\m,\g_\n\}=2~ g_{\m \n},~(\m=0,1,2,3)$,  and $\G_{m n}={1\ov4}
[\G_m,\G_n],$ where $m,~n$ are tangent space indices. 
$\a^a$ are three $4\times 4$ $SU(2)_A$ matrices and along with other three
matrices of $SU(2)_B$ generate the $({1\ov2},{1\ov2})$ representation of the
gauged model\cite{fs}.

\section{ Domain Walls/Membranes}
\subsection{Pure dilaton gravity}
The first solution we obtain is a domain wall  in four dimensions with
the following configuration satisfying the field equations derived from the 
action \eqn{act},
\bea
&&ds^2= U(y)\left( dt^2 - dx_1^2 - dx_2^2\right) - U(y)^{-1} dy^2,\nnu\\
&& \f= -ln~ U, \hskip1cm U=  m |y-y_0|,\nnu\\ 
&& A_\m^a=0,\hskip1cm \eta= 0,
\ll{3}
\eea
where $m^2= \L^2/2$. This background is singular at $y=y_0$ 
which is the position of
the $t-x$ hyperplane or domain wall. 
Since  no matter fields are present other than the 
dilaton this background represents pure dilaton gravity. Note that $e^\f$, 
analog of string coupling, vanishes at asymptotic infinity ($y\to\pm\infty$)
and so also the curvature scalar.
But both are divergent at $y=y_0$. The isometry group for this background is
$P_3(1,2)\times {\cal Z}_2$, where $P_3$ represents three dimensional
P\"oncare group and ${\cal Z}_2$ is the reflection symmetry of the dilaton
potential around $y=y_0$. 

We now  study the supersymmetric properties
of this background. If we substitute \eqn{3} 
in the supersymmetry equations \eqn{2} we find that the fermionic variations
vanish  provided 
the supersymmetry parameters satisfy 

(a) when $e_B=0,~\L=e_A$  
\bea
&&~\bar\ep^I~ \G_3 = i~ \bar\ep^I, \hskip1cm {\rm for}~~ 
U= m |y-y_0|,\nnu\\
&&\bar\ep^I = U^{1\ov 4}~ \bar\ep_0^I
\ll{4}
\eea

(b) and when $e_A=0,~\L=e_B$
\bea
&&~\bar\ep^I~ \G_3\g_5 = -~ \bar\ep^I, \nnu\\
&&\bar\ep^I = U^{1\ov 4}~ \bar\ep_0^I
\ll{4a}
\eea
where $\ep_0^I$ is a constant spinor. The conditions \eqn{4} and \eqn{4a}
can be satisfied because for spacial indices all $\G$'s have imaginary 
eigen values $\pm i$, while $\g_5^2=1$. 
These conditions break half of the supersymmetries in either
case. Thus we see that there exist nontrivial Killing 
spinors preserving $N=2$ supersymmetry for pure dilatonic domain 
wall backgrounds which is sufficient for the stability
of the solution against quantum fluctuations. 

\subsection{ Pure axionic gravity} 

A nontrivial axionic field configuration can  
also be obtained in the vacuum of this theory. 
Let us first  present a much familiar geometry which is the solution of the
field equations derived from (1),
\bea
&&ds^2= \left( (1+Q r^2)dt^2 - {dr^2\ov (1 +Q r^2)} 
- r^2d\theta^2\right) -  dy^2,\nnu\\
&& \f= 0,  \hskip1cm A_\m^a=0,\br 
&& \eta= \pm 2\sqrt{Q}~ y ,
\ll{5a}
\eea
such that $ 4Q = \L^2/2$.  
The geometry of the four manifold is  an anti-de Sitter (AdS) space 
$AdS_3\times R$ ~\footnote{ When y-coordinate is a closed circle the geometry
will be $AdS_3\times S^1$.}. 
This solution is  analogous to the Freedman-Gibbons (FG)
pure electro-vac solution \cite{fg} which has the geometry of $AdS_2\times R^2$,
 we will discuss it in the next section.
 We expect this background to be stable if the stability analysis 
is done based on the methods of anti-de Sitter spaces 
\cite{ff}. Instead we try to  show here   
that the solution \eqn{5a} will be stable from the supersymmetry arguments. 
We present another solution below which has similar asymptotic properties as
the background in \eqn{5a}. 
We write down a new axionic gravity vacuum solution of (1)
\bea
&&ds^2= \left( f^2 dt^2 - {dr^2\ov f^2} 
- f^2 d\theta^2\right) -  dy^2,\nnu\\
&& \f= 0, \hskip1cm f(r)=1+\sqrt{Q}~r \nnu\\ 
&& A_\m^a=0,\hskip1cm \eta= 1\pm 2\sqrt{Q}~ y ,
\ll{5b}
\eea
with $ 4 Q= \L^2/2$ where $2\sqrt{Q}$ is the measure of the axion charge. 
Note that for $Q=1$ the asymptotic properties of both the solutions
\eqn{5a} and \eqn{5b} are the same. But the background in \eqn{5b} is much
transparent from the supersymmetry point of view.
If we substitute \eqn{5b} in eq.(2) we find that the 
background  preserves $1/4$ of
the supersymmetries. Corresponding Killing spinors are  for $e_A=0$
\bea
&&\ep^I = f(r)^{1\ov 2}~ \ep_0^I \nnu\\
&&~\bar\ep^I~ \G_3 = \pm i~ \bar\ep^I, \hskip1cm ~\bar\ep^I~\G_1\g_5=
-\bar\ep^I. 
\ll{5c}
\eea
The twin conditions in \eqn{5c} on the Killing spinors 
break the supersymmetry to $1/4$th. Thus pure axionic gravity in the gauged
model has $N=1$ supersymmetry intact. Other conditions 
when $e_B=0,~\L=e_A$ can also be similarly  derived.

\subsection{ Nonvanishing dilaton and axion}
Now we generalise above two special cases and show that both dilaton 
and axion can exist together in the vacuum of the gauged supergravity
in a supersymmetric fashion.
We obtain correspondingly a solution 
\bea
&&ds^2= U(y)\left( f^2 dt^2 - {dr^2\ov f^2} 
- f^2 d\theta^2\right) - U(y)^{-1} dy^2,\nnu\\
&& \f= -ln~U, \hskip1cm U(y)=1 + m |y-y_0|, \nnu\\
&&\eta= 1+ 2\sqrt{Q}~ y ,\hskip1cm f(r)=1+\sqrt{Q}~r \nnu\\ 
&& A_\m^a=0,\hskip1cm 
\ll{6a}
\eea
such that $ m^2= - 4Q + \L^2/2$.  This solution is regular at $y=y_0$ as
compared to domain wall solution in \eqn{3}\footnote{We could have as well
chosen $U=1+ m~|y-y_0|$  in eq.\eqn{3} instead of taking a divergent solution
at $y=y_0$.}. 
The geometry  of the manifold is $AdS_3\times R$. 
Thus the Minkowskian  flat geometry in (3) has turned into an anti-deSitter  
geometry in presence of nontrivial axion flux. 
The corresponding isometry group
now is $O(2,2)$ of the $AdS_3$, 
$Z_2$ symmetry is broken due to nontrivial axion field. 
This background preserves $1/4$-th of the supersymmetries. 
We find that most general Killing spinors can be written when 
$m^2 = e_A^2/2$ and $4Q= e_B^2/2$,
\bea
&&\ep^I =  U(y)^{1 \ov 4}~f(r)^{1\ov 2}~ \ep_0^I \nnu\\
&&~\bar\ep^I~ \G_3 = i~ \bar\ep^I, \hskip1cm ~\bar\ep^I~\G_1\g_5=
-\bar\ep^I. 
\ll{6b}
\eea
It is clear from \eqn{6a} that pure dilatonic and pure axionic solutions are
special cases of \eqn{6a} when $Q=0$ and  $m=0$ respectively.
Note that in \eqn{6a} only positive values for the axionic charges $\sqrt{Q}$
 are allowed.
This restriction comes from the supersymmetry in presence of a regular
dilaton field 
which is nontrivial now. While in the previous case of pure axionic gravity 
(no dilaton) both plus and minus values for the axion charge were allowed.

\section{ Vortices/Strings}

\subsection{Pure dilatonic Case}
We also  obtain  vortex or string-like solution of the  equations of motion 
derived from action \eqn{act}. We look for the solutions such that the
isometry group is $P_2(1,1)\times SO(2)$.  
Again we consider pure dilaton gravity.
The background solution is 
\bea
&&ds^2= U(y)^2\left( dt^2 - dx^2 \right) - 
 dy^2 - y^2~d\theta^2,\nnu\\
&& \f= -2~ln~ U, \hskip1cm U=  m ~y, \nnu\\ 
&& A_\m^a=0,\hskip1cm \eta= 0,
\ll{6}
\eea
where the parameter $m$ satisfies $ m^2= \L^2/8$. 
Here $y=\sqrt{y_1^2+y_2^2}$ is the radial distance
in the transverse 2-dimensional Euclidean 
plane and $\theta$ is the azimuthal angle. 
It should be noted that the function $U(y)$ in \eqn{6} is of very
specific nature which vanishes at $y=0$. It indicates that the volume of the 
$t-x$ space shrinks to zero at $y=0$ much like the size of the 2-sphere 
shrinking to zero at the centre of the Cartisian three-space. 
For this background the curvature scalar 
is singular at the origin $y=0$ and also the string
coupling while at asymptotic infinity both are vanishing. 

Background in \eqn{6} preserves half 
of the supersymmetries of the theory. There exist a nontrivial Killing spinor
for $e_B=0$  
\bea
&& \bar\ep^I~\G_2 = i~ \bar\ep^I, \nnu\\
&&\ep^I = U^{1\ov 2}~ \ep_0^I,\hskip1cm {\rm for }~~ U=m~y.
\ll{7}
\eea
where $\G_2$ corresponds to the direction tangent to y.
Note that these Killing spinors vanish at $y=0$ .
Other Killing spinors when $e_A=0$ can also similarly be derived.

\subsection{ Presence of nonvanishing electric field} 

 As in the case of Domain-walls, we find here that a nontrivial electric flux 
can be switched on in the supersymmetric string solution \eqn{6}. 
The electric field is given by a constant electric field tangent to 
the x-direction. Before showing 
that we assume that the singularity at the origin is modified in presence of
electric field. We consider that at the origin $U~|_{y=0}=1$  so that
string coupling is unity at the origin. Advantage of this is that we can take 
the smooth limit $m\to 0$ in the background below. The limit correponds to 
constant dilaton field. It will be clear soon. 
Under these  specifications the background solution of action (1) in presence 
of electric field is
\bea
&&ds^2= U(y)\left( (1+ {q^2\ov 2} x^2)dt^2 - {dx^2\ov (1+{q^2\ov 2}
 x^2)} \right) - dy_1^2 - dy_2^2,\nnu\\
&& \f= -~ln~ U, \hskip1cm U= 1+ m~ y^2, \nnu\\ 
&& A_\m^a=(q~ x, 0, 0, 0)~\delta^{a 3},~ \eta= 0,
\ll{8}
\eea
such that  $ m={\L^2-q^2\ov 8}$, where $q$ is the measure of the electric field
strength \footnote{ When we take $q\to 0$ the background \eqn{8} is still a
vacuum solution}.
From \eqn{8} it is clear that we have chosen an
specific direction in the isospin space thus breaking $SU(2)_A$ to 
$U(1)$. Note that now there is no singularity at $y=0$ and asymptotically 
string coupling and curvature scalar both vanish. 
Though  background \eqn{8} does not preserve
any supersymmetry but we are free to consider the limit 
$m\to 0, ~q\to\L,~e_A=0$. This is the advantage of choosing the potential 
$U=1 + m~r^2$. Now we can set $m=0$ in \eqn{8}. The vacuum solution in \eqn{8} 
then reduces to Freedman-Gibbons  purely electric `electro-vac' solutions. 
It is 
easy to follow this by defining the following relations
\be
x= {1\ov q} \tan\r,\hskip1cm t={\tau\ov q},~~~ ~~~-{\pi\ov 2}\le \r\le{\pi\ov2}.
\ll{9}
\ee
Eq. \eqn{8} becomes,
\bea
&&ds^2= {1\ov \L^2 \cos^2\r}\left( d\tau^2 - d\r^2 \right) - 
 dy_1^2 - dy_2^2,\nnu\\
&& \f= 0, \hskip1cm U= 1, \nnu\\ 
&& A_\m^a=(q~ x, 0, 0, 0)~\delta^{a 3},~ \eta= 0,
\ll{10}
\eea
which is the electro-vac solution of \cite{fg} corresponding to case $e_A=0, 
B_\m^a=0$. 
This background has $N=2$ supersymmetry \cite{fg} 
and can be described as the extremal (or supersymmetric) limit $m\to 0$  
of the background in \eqn{8}. 

\subsection{ Presence of axion charges}

In the presence of axion field we obtain a vacuum solution for the action (1) 
which is a generalisation of the 
fundamental string solution \cite{dh}. Such solutions (non-static) with an
isometry group $O(1,1)\times SO(2)$ have already been obtained 
in  a previous work \cite{hs} by the author. But here we are going to
present the background solution which is static and has the symmetry 
$R^1\times Z_2\times SO(2)$. This background is 
\bea
&&ds^2= \left( f(x) dt^2 - {dx^2\ov f(x)}\right) - U(y,x)\left(dy^2 + 
y^2~d\theta^2\right),\br
&& \f=-ln~U(y,x),\hskip1cm U=1 - Q~ln~|y-y_0| + m~|x-x_0|,\br
&& f(x)= 1 + m ~|x-x_0|,\hskip1cm \pa\eta= ^{*} e^{-2\f} H,\br
&& B_{01}= U^{-1}
\ll{10aa}
\eea
where $H$ represents antisymmetric field strength of $B_{\m\n}$
and $ m^2=\L^2/2$. Note that this background
has invariances under time translations,  reflections about $x=x_0$ plane and
rotations in the two dimensional y-plane. This background, however, does not
preserve supersymmetries in (2). But it has two special limits; one when $Q=0$
\eqn{10aa} reduces to the supersymmetric domain wall solution in \eqn{3}, other
when $m=0$ background \eqn{10aa} becomes a fundamental string \cite{dh}. 
Thus we claim that the background \eqn{10aa} is the most simple generalisation
of string geometry, though unstable, in presence of dilatonic potential 
(cosmological constant) as in (1). That is to say, background \eqn{10aa}
represents a
fundamental string whose vacuum is a domain-wall instead of a 
Minkowski space. The string vacuum (domain-wall) is stable but not
the strings (geometric excitations which carry axionic charges) in this
vacuum.
The stable geometries in the domain-wall vacuum which carry axionic  
charges have already been presented in section III.

\section{ Point-like Solutions}
 To complete the analysis we also obtain maximally symmetric isotropic 
background solutions of \eqn{act}. We have not been able to find a 
supersymmetric case for them. The solution we are going to present below 
is  of very specific type that it allows only unit value of the 
magnetic charge. The solution is  given by
\bea
&& ds^2= f~ dt^2 -{dr^2\ov f} - U~\left(d\theta^2 + 
\sin^2{\theta}~ d\F^2\right),\nnu\\
&& \f= -ln~ U , \hskip1cm U= 1+ m~ r, ~~~f(r)= {r\ov r_0}\pm 1,  \nnu\\
&& A_\m^a=(0,0,0,Q_m(1-\cos\theta))~\delta^{a3}, ~~~\eta=0
\ll{12}
\eea
with $m={\L^2+Q_m^2\ov2}~r_0$ such that $Q_m=1$. Thus this solution exists
only in the presence  unit value of magnetic charge.  For plus sign in $f(r)$
the solution is smooth every where and is asymptotically flat. But for minus
sign there appears a nonsingular horizon at $r=r_0$. From the nonsingular 
horizon we mean by the finite size. However, this does not hide any
singularity. 

\section{Summary}

We summarise our results as follows:

	In this work we have  obtained a  stable domain-wall
solution in the purely dilatonic vacuum of gauged $N=4$ supergravity. 
The solution preserves half of the supersymmetries. 
Then we obtained pure axionic
gravity vacuum which preserves $1/4$ of the supersymmetries of the theory.  
We also have obtained domain-walls in the presence of nontrivial axionic
background. To our surprise these backgrounds also preserve one fourth
of the supersymmetries and are therefore stable. As we find that the domain
wall solution cannot be given an interpretation of a dual of a 4-form
field strength in the action (1) because the dilatonic potential is unbounded
from below.  
Though, in the case of type IIA gauged supergravity\cite{roma}
the domain  walls have an interpretation as a dual of a 10-form 
field strength of an 8-D-brane.  

Next, we have obtained vortex (string-like) solutions which preserve 
half of the supersymmetries and therefore are stable solutions. Then we have  
generalised these solutions in presence of constant electric flux  and showed
that these solutions are generalisation 
of Freedman-Gibbons electro-vac background. 
Though these solutions do not preserve any supersymmetry 
but can be smoothly taken to their extremal (supersymmetric) 
limit which is  an electro-vac background preserving $N=2$ supersymmetries. 
 
In last we obtained the maximally symmetric point-like solutions also. We
have'nt found any fraction of supersymmetry being left unbroken for point
case. It should be explored further if black holes could be found in
gauged $SU(2)\times SU(2)$ supergravity. 

In conclusion we have seen that the vacuum of the gauged supergravity is 
quite interesting and is rich with  stable brane-like configurations. 

\vskip1.5cm
\noindent{\bf Note Added:} When this work was communicated I came to know
from  Prof. P. K. Townsend that another domain-wall solution 
has been obtained in \cite{co}.  We can see that this domain-wall is related to 
the domain-wall solution
in \eqn{3} by going to a frame where the potential $U(y)\to 1/U(y)$.
However, the  string coupling  diverges at 
asymptotic infinity in that case. One thing can also be seen from \eqn{3} that
the dilaton potential $-\L^2/2 e^\f$ is bounded because $e^\f \to 0$ as 
$y\to \pm\infty$.

\noindent{\bf Acknowledgements:} I would like to thank  Mario Tonin for
many useful discussions. I  also thank anonymous referee for bringing
up the previous works  \cite{abs,afk}.

\end{document}